\documentclass[a4paper,11pt]{article}
\pdfoutput=1 

\usepackage{jcappub} 
\usepackage{aas_macros} 

\usepackage[T1]{fontenc} 
\usepackage{mathtools}
\newcommand{\degree}{\ensuremath{\mathrm{^\circ}}}
\newcommand{\cm}{\ensuremath{\,\textrm{cm}}}

\newcommand{\msun}{\ensuremath{\textrm{M}_\odot}}
\newcommand{\gaia}{\textsl{Gaia}}
\newcommand{\fermi}{\textsl{Fermi}}
\usepackage{array}
\newcolumntype{P}[1]{>{\centering\arraybackslash}p{#1}}
\newcolumntype{M}[1]{>{\centering\arraybackslash}m{#1}}
\usepackage{graphicx}
\usepackage[dvipsnames]{xcolor}
\usepackage{longtable}
\usepackage{url}
\usepackage{verbatim}
\usepackage[utf8]{inputenc}
\usepackage{makecell}
\usepackage{comment}
\usepackage{cleveref}
\crefname{section}{§}{§§}
\usepackage[T1]{fontenc} 
\graphicspath{ {Figures/} }

\title{\boldmath Machine-Learned Dark Matter Subhalo Candidates in the 4FGL-DR2: Search for the Perturber of the GD-1  Stream}

\author[a,b,c]{Nestor Mirabal}
\author[d]{Ana Bonaca}

\affiliation[a]{Mail Code 661, Astroparticle Physics Laboratory,
NASA Goddard Space Flight Center, Greenbelt, MD 20771,
USA}
\affiliation[b]{University of Maryland, Baltimore County, MD 21250, USA}
\affiliation[c]{Center for Research and Exploration in Space Science and Technology, NASA Goddard Space Flight Center, Greenbelt, MD 20771}
\affiliation[d]{Center for Astrophysics | Harvard \& Smithsonian, 60 Garden Street, Cambridge, MA 02138, USA}

\emailAdd{nestor.r.mirabalbarrios@nasa.gov}
\emailAdd{ana.bonaca@cfa.harvard.edu}

\abstract{
The detection of dark matter subhalos without a stellar component in the Galactic halo remains 
a challenge. 
We use supervised machine learning to identify high-latitude
gamma-ray sources with dark matter-like spectra
among unassociated gamma-ray sources in the 4FGL-DR2. Out
of 843 4FGL-DR2 unassociated sources at  $|b| \geq 10\degree$, we select 73 
dark matter subhalo candidates. Of the 69 covered by the {\it Neil Gehrels  Swift
Observatory (Swift)}, 17 show at least one X-ray source within the 95\% LAT error ellipse and 52 where we identify 
no new sources.  
This latest inventory of dark subhalos candidates
allows us to investigate
the possible dark matter substructure responsible for the perturbation
in the GD-1 stellar stream. 
In particular, we examine the possibility that the alleged GD-1 dark subhalo 
may appear as a 4FGL-DR2 gamma-ray source from dark matter 
annihilation into Standard Model particles.
}
\begin{document}
\maketitle
\flushbottom

\section{Introduction}
\label{sec:intro}

It has been long hypothesized that the Milky Way halo is populated by   thousands of dark matter subhalos above a mass of $10^6 {\rm M_\odot}$ \cite{Klypin,diemand,springel}. 
With dedicated optical surveys, the number of detected dwarf galaxies
orbiting the Milky Way has grown significantly and sets a
lower bound on the abundance of dark matter subhalos \cite{simon}.   
However, there might be dark substructures that are too faint or
too distant to be picked up by optical surveys alone \cite{Zavala2019}. 
Localizing dark matter subhalos without detectable
stellar populations 
is still of critical
importance not only to understand structure formation but also to 
reveal the particle nature of dark matter directly. 

Weakly interacting massive particles (WIMPs) 
in the nearest and most massive subhalos 
could produce observable gamma-ray sources with 
significantly curved energy spectra 
from annihilating dark matter \cite{Baltz}.
In general, one can parametrize WIMP annihilation spectra with 
a super-exponential cutoff power law \cite{Calore17,Coronado_Blazquez2019}:

\begin{equation}\label{eq:parametric_spectra}
\frac{dN_{DM}}{dE}\left(E\right)=K\left(\frac{E}{E_{0}}\right)^{-\Gamma}e^{-\left(\frac{E}{E_{cut}}\right)^\beta},
\end{equation}

\noindent where $K$ is a prefactor, $E_{0}=10^3$ MeV is the pivot energy, 
$\Gamma$ is the spectral photon index, 
$E_{cut}$ is the cutoff energy and $\beta$ is the curvature index. 
Taking advantage of the extraordinary map of the gamma-ray sky produced by the
 Large Area Telescope (LAT) on
the {\it Fermi Gamma-ray Space Telescope} ({\it Fermi}),
several searches have used source catalogs released by
the LAT collaboration to look for this distinctive spectral 
feature among unassociated sources
\cite{BuckleyHooper10,fermi_dm_satellites_paper,nieto,Mirabal2012,Belikov2012,Zechlin12,BerlinHooper14,Bertoni15,Bertoni16,Mirabal2016,Saz,Calore17,Coronado_Blazquez2019}.
Although there are interesting subhalo candidates with
exponential cutoffs, it is still not yet possible to distinguish them from gamma-ray pulsars \cite{Bertoni15}.

An alternate approach is based on the idea that dark matter subhalos orbiting in the
Milky Way halo might induce gaps when they cross elongated
dynamically-cold 
stellar streams \cite{ibata,johnston,carlberg, yoon,carlberg2012}. 
With astronomical surveys such as the {\it Gaia} mission, there is now sufficient precision to study stellar streams in extreme detail. The GD-1 stellar stream is currently the most notable
example of a possible stream crossing \citep{carlberg16,bonaca}.
The observed morphology of GD-1 (gap and off-stream spur of stars) is naturally reproduced in models of the stellar 
stream that include an encounter with a massive object
($\sim 10^6-10^8$ M$_\odot$), like a globular cluster, a dwarf galaxy, or a dark matter subhalo \citep{bonaca, deboer2020}.

Each of these techniques offers excellent opportunities 
to survey the subhalo population.
But a combination of approaches might boost our chances for success. 
In this paper, we use the first incremental version of the Fourth {\it Fermi}-LAT
Catalog of gamma ray sources (4FGL-DR2, for Data Release 2) to
select possible dark matter subhalos using machine-learning
techniques \citep{Mirabal2012,Mirabal2016}. Out of
843 unassociated gamma-ray sources at $|b| \geq 10^{\circ}$, we 
extract a subset of 73 sources that are consistent with
the spectral shape of annihilating dark matter. We then
search for X-ray sources in the LAT 95\% error ellipses using 
archival {\it Swift} data. We next compare the 
locations of the subhalo candidates with 
 the predicted locations of the GD1's perturbing subhalo from {\it Gaia}.   

\section{Dark Subhalo Selection}
\subsection{Dataset}
We start with the 4FGL-DR2 covering the time period August 4, 2008, to 
August 2, 2018 in the 50 MeV-1 TeV energy range 
\cite{4FGL,4FGLDR2}. 
For our study, we use the
latest version available at the time of
this writing \texttt{gll-psc-v27.fit}. These 
sources are extremely well characterized by 74 catalog columns. 
Of the 5787 4FGL-DR2 sources, 1670 are unassociated. This leaves
4117 associated sources. Since we are only interested in searching for dark 
subhalo candidates away 
from the Galactic plane, we can safely remove Galactic plane sources 
such as supernova remnants and pulsar wind 
nebulae. The largest source population  at 
high Galactic latitude is extragalactic. We parametrize these sources with 
an extragalactic set that includes 3505 training sources  
from all AGN classes in the 4FGL. The extragalactic set includes
11 non-blazar active galaxies (AGN,agn)  
1382 blazar candidates of uncertain type (BCU,bcu), 1308 BL Lacs (BLL,bll), 
743 flat-spectrum radio quasars (FSRQ,fsrq), 
9 narrow-line Seyferts 1 (NLSY1,nlsy1), 
44 radio galaxies (RDG,rdg), 2 soft spectrum radio quasars (ssrq), 
5 compact steep spectrum quasar (css), and 1 Seyfert galaxy (sey). 

In order to generate a model for 
dark subhalo searches we harness the vast catalog of 
gamma-ray pulsar and globular cluster spectra \cite{Mirabal2016}.
Early studies of unassociated gamma-ray sources led to 
the important realization that gamma-ray
pulsar spectra would be nearly indistinguishable from the super-exponential 
spectra generated by WIMP
annihilation in dark subhalos
\citep{Baltz}.
Thus formally, a high-latitude, non-variable
gamma-ray pulsar candidate without detected pulsation is also a potential
subhalo candidate. For our subhalo set, we include 
235 identified pulsars (PSRs), 30 globular clusters (glc), and 24 
associated pulsars (psrs)
with spectra that are also consistent with 
annihilations of $\sim$ 20--70 GeV dark
matter particles \cite{Bertoni15,Bertoni16}. 

We end up with a grand total of 3794 training sources. 
With 3505 extragalactic and 289
pulsars, our training
set is imbalanced. Whenever there is 
imbalance, machine learning classification tends to be
 artificially skewed  
towards the majority class. 
As in \cite{Mirabal2016}, we use the Synthetic Minority Over-sampling
Technique to balance the class distribution 
(SMOTE, \cite{Chawla}). SMOTE adds synthetic pulsars sampled from the 
existing pulsars 
using $k$-nearest neighbors until they both contain near equal samples. 

\subsection{Machine Learning Algorithms}
For 4FGL-DR2 training, testing  and classification,
we rely on the scheme introduced in \cite{Mirabal2016}.
For training we use Random Forest and eXtreme Gradient Boosting (XGBoost) 
tree algorithms.
Although both rely on classification trees, Random Forest uses feature bagging 
that makes
decisions by a committee of individual classification trees.
Each tree casts a vote for the predicted class and the majority vote is taken 
as the prediction,
thus it reduces variance
\citep{Breiman}.
XGBoost \footnote{https://github.com/dmlc/xgboost}  
tries to do better at each interaction, thus it reduces
bias \citep{Chen}. We settle on a combination of R and {\bf scikit-learn} 
\footnote{http://www.scikit-learn.org} for the rest of the paper.

\subsection{Feature Selection}
\begin{figure}[!t]
\centering
\includegraphics[height=4.3cm]{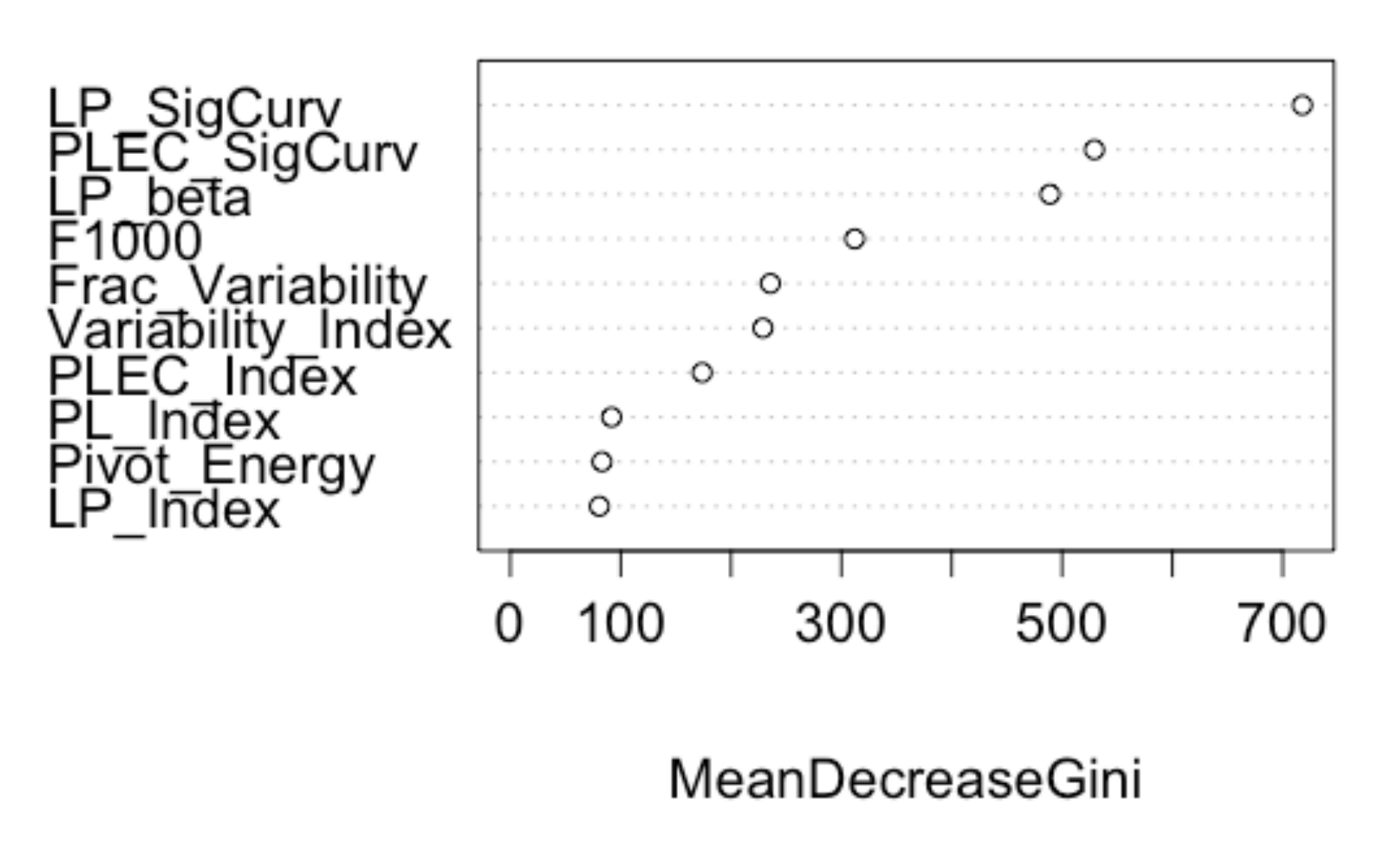}
\includegraphics[height=4.1cm]{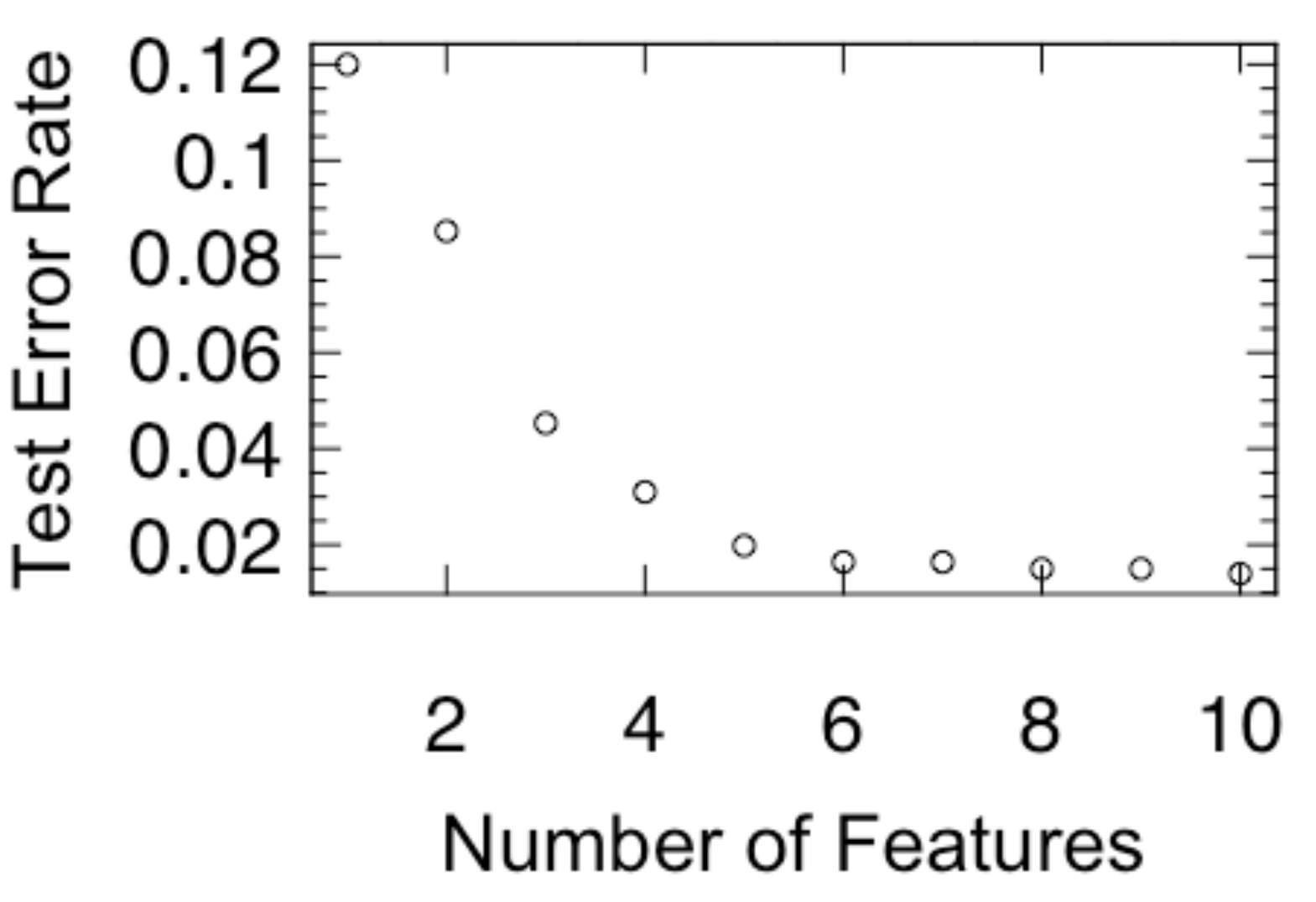}
\caption{Left: Importance of the 10 top features that were used to train the Random Forest classifier. 
Right: Out-of-bag performance as a function of the number of 
features used by 
the
classifier.
}
\label{fig1}
\end{figure}

Given the large number of 4FGL-DR2 features, it is critical to  
determine their relative importance for 
correct classification. To do this we rely on the
Gini coefficient as determined by Random Forest \cite{gini,liaw}. 
A larger value of Gini importance means that a 
particular feature plays a greater role in predicting a certain class.
This is shown in Figure~\ref{fig1}. As shown,  
the most important predictors are:
\begin{itemize}
\item  \texttt{LP\_SigCurv:} Significance (in $\sigma$ units) 
of the fit improvement between PowerLaw and
LogParabola. 
\item  \texttt{PLEC\_SigCurv:} Same as LP\_SigCurv for PLSuperExpCutoff model
\item \texttt{LP\_beta:} Curvature parameter when fitting with LogParabola
\item \texttt{Flux1000:} Integral photon flux from 1 to 100 GeV
\item \texttt{Frac\_Variability:} Fractional variability computed from the fluxes in each year
\item \texttt{Variability\_Index:} Sum of 2$\times$log(Likelihood) difference 
between the flux fitted in each time interval and the average flux over the 
full catalog interval
\item \texttt{PLEC\_Index:} Low-energy photon index 
when fitting with PLSuperExpCutof
\item \texttt{PL\_Index:} Photon index when fitting with PowerLaw
\item \texttt{Pivot\_Energy:} Energy at which error on differential flux is 
minimal
\item \texttt{LP\_Index:} Photon index at Pivot Energy 
when fitting with LogParabola
\end{itemize}

Another
reliable importance measure is the evaluation of ``out-of-bag'' (OOB) performance as a 
function of  number of features. As can be seen Figure \ref{fig1}, 
the OOB performance tends to settle 
after 8 or more features 
are included. Since there is no obvious learning loss, we decided to retain the top 10 features for the rest of the 
classification. 

\subsection{Training and Classification}
\begin{figure}[!t]
\centering
\includegraphics[height=5cm]{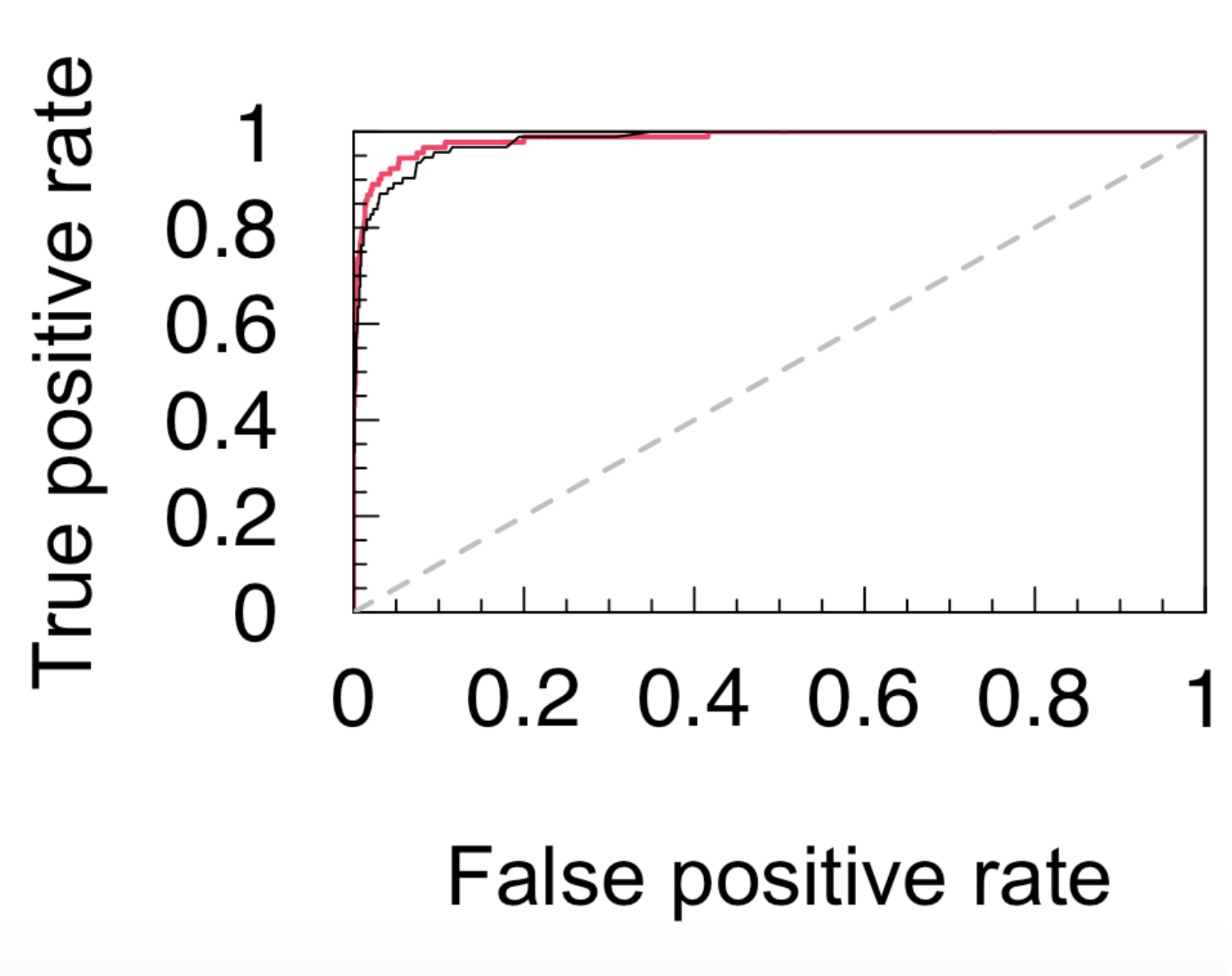}
\caption{ROC curve comparison of the two methods. The red thick line shows
XGBoost while the black thin line shows the curve for Random Forest.}
\label{fig2}
\end{figure}

For cross validation, we divide the  
dataset into training (70\%) and
testing (30\%) subsets. Overall, the Random Forest classifier achieves an  accuracy of 
97.7\%. Similarly, XGboost reaches an accuracy of 97.4\%.
The fluidity of ongoing source association efforts offers another kind of sanity check. 
As our work evolved
from the initial \texttt{gll-psc-v23.fit} 4FGL-DR2 file release to the current 
\texttt{gll-psc-v27.fit} version\footnote{ https://fermi.gsfc.nasa.gov/ssc/data/access/lat/10yr\_catalog/}, 
five of our machine-learned subhalo candidates moved from unassociated to 
pulsar associations\footnote{4FGL J0312.1$-$0921,4FGL J1221.4-0634,
4FGL J1304.4+1203, 4FGL J1400.6-1432, 4FGL J2039.4$-$3616} and are no
longer included here.  
Figure \ref{fig2} compares the Receiver Operating Characteristic curves (ROC
curve) for the two methods.  ROC curves allows us to visualize the True Positive Rate (sensitivity) against False Positive Rate (specificity). Curves closer to the top-left corner indicate a better performance.

\begin{figure}[!t]
\centering
\includegraphics[width=0.95\textwidth]{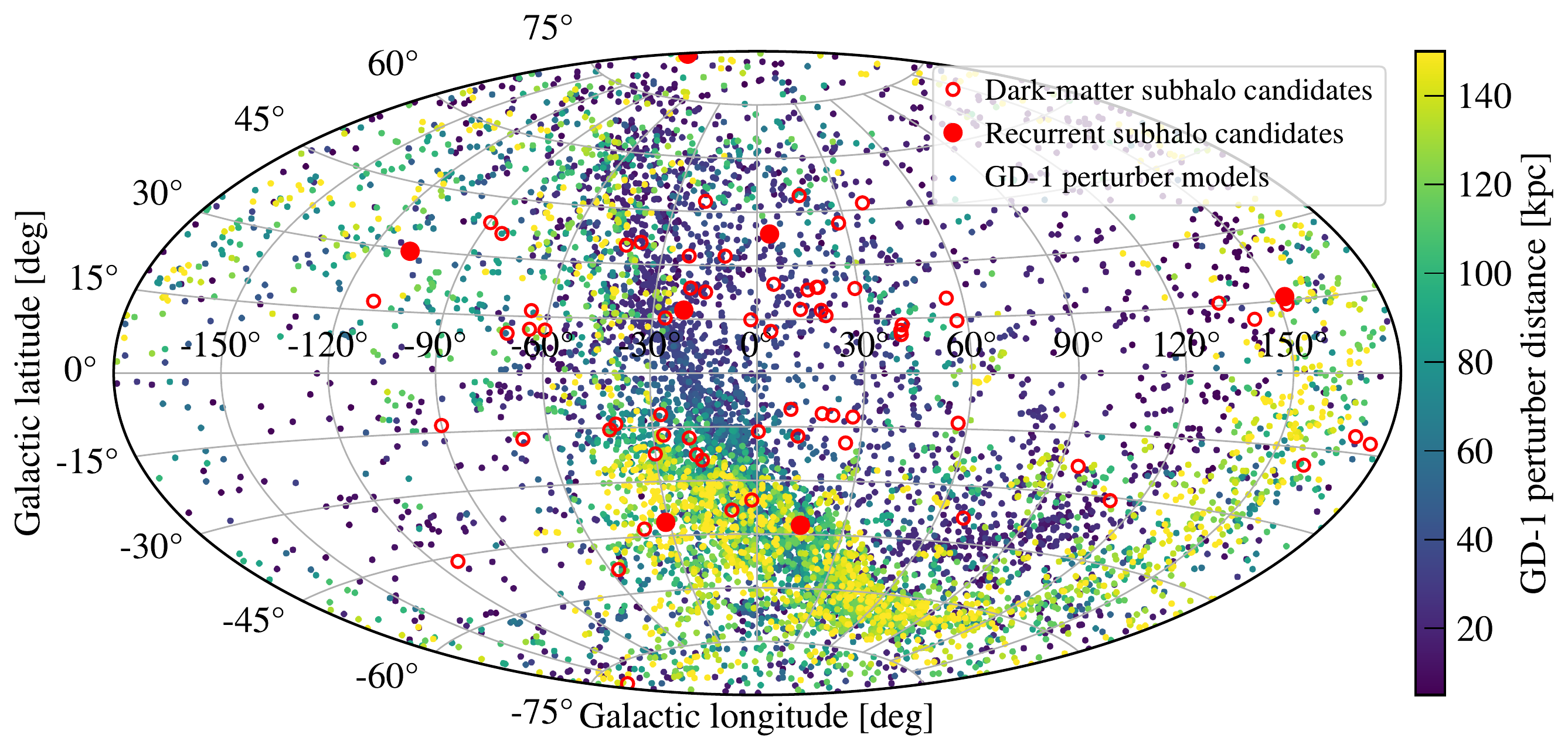}
\caption{Locations of 73 dark subhalo candidates in the Galactic sky coordinates (empty red circles).
Candidates recurring in multiple \fermi\ data releases are marked as solid red circles.
The color-coded dots indicate the predicted distance of the GD-1 perturbing subhalo.}
\label{fig3}
\end{figure}

Having trained the classifiers using labeled LAT sources, we can now identify dark subhalo candidates
among unassociated 4FGL-DR2 sources. 
To cut back on contamination from Galactic plane objects, 
we only use unassociated 
sources at high latitude 
($|b| \geq 10^{\circ}$). After data cleaning, we
end up with 843 unassociated sources at $|b| \geq 10^{\circ}$. 
Applying our machine-learning
models to this sample,
we end up with 73 4FGL-DR2 sources  which both our methods predict to be dark subhalos. Table \ref{table1} presents
the list of sources consistent with being subhalo candidates. Figure \ref{fig3} shows their location in RA-Dec (top) and 
Galactic (bottom) coordinates.

\section{X-ray Observations}
For each source in Table \ref{table1}, 
we select all observations available in 
the {\it Swift} archive.
The majority are part of a long-term follow-up program of
unassociated {\it Fermi} sources \cite{stroh}.
In total, we find that 69 of the 73 sources have had their 95\% error ellipse
covered. For the 
 X-ray Telescope (XRT) analysis, we used  XRTDAS\footnote{https://swift.gsfc.nasa.gov/analysis/xrt\_swguide\_v1\_2.pdf} within HEASOFT 6.26. 
We used the standard selections of
$0-12$ in  the Photon Counting (PC) mode and performed the analyses 
in the 0.3--10 keV energy range. In case of a point source detection within
the 95\% error ellipse, we extract the number of photons from 
a circular region around the source with
radius 20 arcseconds and a 40-150 arcsecond annular region  to describe the 
background and scale it to the source region. In total,
17 4FGL-DR2 sources show one or more X-ray detections within their 95\% 
error ellipse. Whenever there is a detection, 
 X-ray fluxes are
calculated assuming photon index 2 and the full Galactic 
$N_{\rm H}$ estimated using WebPIMMS
\footnote{https://heasarc.gsfc.nasa.gov/cgi-bin/Tools/w3pimms/w3pimms.pl}. 
See Table \ref{table2} for flux values. 

In order to place X-ray upper limits, we followed a procedure similar to
\cite{heinke}.
We extract the number of photons from the brightest 
circular region within the LAT 95\% ellipse. Using annular
background regions, 
we compute the expected number of background photons and
compute the 90\% confidence lower limit in the background region 
\cite{gehrels}. The upper limit is
derived by  subtracting this lower limit 
from the observed photons in the source 
region. 
To determine upper limits for nondetections, we
assume an
absorbed
power law with photon index 2 and $N_{\rm H}$ from
WebPIMMS. Results are listed in Table \ref{table1}. 
In order to double check our results, we
used the online {\it Swift} XRT products generator\footnote{https://www.swift.ac.uk/} 
based on tools developed for the 
2SXPS Swift XRT Point Source Catalogue \cite{evans}.
Finally, we also consulted the online version of
the {\it Swift}-XRT Survey of Fermi Unassociated Sources program
for consistency
\footnote{https://www.swift.psu.edu/unassociated/}.

\section{Searching for the GD-1 Perturber}
Searches for low-mass dark-matter subhalos in the Milky Way galaxy focus on detecting their effects on the motions of halo stars.
Stellar streams are especially sensitive tracers of such low-amplitude gravitational perturbations \citep{johnston2002, ibata2002}.
Formed by tidal dissolution of a progenitor star cluster or dwarf galaxy, stellar streams are long, thin and kinematically cold groups of stars moving on similar orbits through the Galaxy for billions of years \citep{odenkirchen2001, rockosi2002, kupper2010}.
A subhalo passing by a stellar stream imparts measurable velocity kicks to its member stars. 
Less massive subhalos merely increase the stream's velocity dispersion \citep{johnston2002}, while more massive ones significantly alter stellar orbits to produce an underdensity, or a ``gap", in the stellar stream \citep{yoon}.
Parameters of the impact, including the subhalo mass and time of impact, remain recorded in the density and kinematic structure of the perturbed region \citep{erkal2015, erkal2016}.
Early analyses of photometric data from the Sloan Digital Sky Survey (SDSS) revealed an abundance of stream gaps consistent with perturbation by a population of CDM subhalos \citep{carlberg2012b, carlberg2013}.
However, deeper photometry indicates that at least some of these density variations may be due to contamination from other Milky Way stars \citep{ibata2016}, rendering the detection of dark-matter subhalos with streams inconclusive in SDSS.

Proper motions provided by the \gaia\ mission vastly improved the selection of stream member stars.
Applied to the GD-1 stellar stream, this selection revealed significant gaps in the stream density, as well as a spur of stream stars beyond the main stream track \citep{pwb}.
Numerical experiments show that a recent ($\approx0.5\,$Gyr) impact of a massive ($10^6\,\msun - 10^8\,\msun$) and compact ($\lesssim20\,$pc) object can produce features observed in GD-1 \citep{bonaca}.
Even though a spur-like structure can be induced in GD-1 by the Sagittarius dwarf galaxy \citep{deboer2020}, orbital integrations in the fiducial model of the Milky Way show that none of the known objects like dwarf galaxies, globular clusters, and molecular clouds, approach GD-1 sufficiently close to reproduce the structure of the GD-1 gap-and-spur features in detail \citep{bonaca}.
This allows the possibility that GD-1 encountered a low-mass dark-matter subhalo.

The most direct way to ascertain the nature of a stream perturber would be to observe it directly.
Detailed spatial distribution (the location and size of the observed gap and spur features) and kinematics (small radial velocity offset detected between the stream and spur) of the perturbed region of GD-1 constrain the perturber's orbit \citep{bonaca}.
Present-day sky positions of perturbers on allowed orbits are shown in Figure~\ref{fig3} and color-coded by their distance from the Sun (darker colors for more nearby models).
\fermi\ dark-matter subhalo candidates are overplotted as large red circles.
The present-day positions of the GD-1 perturber were calculated assuming an axisymmetric, static, analytic model of the Milky Way's gravitational potential, whereas the most recent dynamical studies suggest that the halo shape is spatially complex and still evolving due to ongoing mergers with Sagittarius and Large Magellanic Cloud \citep{vasiliev2021, garavito-camargo2021}.
Therefore, each individual perturber position is likely associated with a systematic uncertainty stemming from an overly simplified Milky Way model.
Still, positions of a number of subhalo candidates overlap the general area plausibly occupied by the GD-1 perturber and provide valuable targets for follow-up observations.

Interestingly, the distribution of \fermi\ subhalo candidates is asymmetric with respect to the Galactic center:
most candidates are located at Galactic latitudes $|b|\lesssim45\degree$, but with a larger extension to negative Galactic longitudes, $-100\degree\lesssim l \lesssim60\degree$.
Even more strikingly, the preferred sky positions of the GD-1 perturber are along a great circle.
Dynamically, their orbits appear to be aligned with the orbital plane of the Sagittarius dwarf galaxy \citep{bonaca}.
The robustness of these asymmetries needs to be further tested, e.g., with a more comprehensive search of the orbital parameter space in the case of the GD-1 perturber, and with a more detailed accounting of \fermi\ sensitivity to  subhalo signals.
If confirmed, the asymmetric distribution of subhalo candidates suggests that at least a fraction of dark matter in the Milky Way is unrelaxed and that coherent debris flows from recent mergers \citep[e.g.,][]{dierickx2017, besla2019, garavito-camargo2021, naidu2020} need to be accounted for in studies of dark matter on Galactic scales.

\section{Is GD-1's Perturbing Subhalo Detectable in Gamma Rays?}
To start answering this question, we must rely on cosmological simulations and $J$ factor predictions.
The expected gamma-ray flux from dark matter annihilation \cite{ulio} is given by:

\begin{equation}
\frac{d\phi(\Delta \Omega)}{dE_\gamma} = \left( \frac{1}{4\pi}\frac{dN_\gamma}{dE_\gamma}\frac{\langle\sigma v\rangle}{2m_{DM}^2}\right) \left( \int_{\Delta\Omega} d\Omega \int_\text{l.o.s.} d\ell ~\rho^2_\chi(\vec{\ell}\,) \right).
\end{equation}

\noindent The left term within parentheses represents the particle physics contribution where 
 $\langle\sigma v\rangle$ is
the velocity-averaged annihilation cross section, $m_{DM}$ is the WIMP mass,
and $dN_\gamma/dE_\gamma$ is the differential spectrum of gamma rays from
annihilation of a pair of DM particles. The right hand side term (as shown in parentheses) represents the $J$-factor $J$, which encloses all the astrophysical considerations. The first integral is performed over the solid angle  of a region of interest  (ROI, $\Delta\Omega$). 
The second integration is performed along the line of sight, and $\rho(\vec{\ell})$
is the density of the DM particles.

Finding astronomical objects that maximize $J$ is key to achieving a detection.
Using predictions from the Via Lactea II \cite{diemand} dark-matter only simulation, \cite{Coronado_Blazquez2019} find
significantly higher $J$-factors associated to low-mass subhalos. With respect to previous works, 
\cite{Coronado_Blazquez2019} assigns ``proper'' subhalo
concentrations and repopulated the VL-II simulations
with  low-mass subhalos down to $10^3$ M$_\odot$. 
The main limitation in these results
is the unknown impact of baryons in the
subhalo population.

Possibly the most careful job assessing the gamma-ray detectability of dark matter
subhalos taking into account baryons and uncertainties in different cosmological
simulations is by \cite{Calore19}. Using a 
set of four configurations including {\it Aquarius},
{\it Phat-ELVIS}, Stref and Lavalle, \cite{Calore19} find that
there should be anywhere between 4 and 50 subhalos in the
$\sim 10^6-10^8$ M$_\odot$ mass range detectable in the 3FGL
catalog between 15 kpc and 160 kpc. 
Relative to the 3FGL catalog, the 4FGL-DR2 used in this work has twice as much 
exposure. 

Assuming a massive
$\sim 10^6-10^8$ M$_\odot$ perturbing subhalo (as in the previous section) and
a 100 GeV dark matter particle annihilating to $b\bar{b}$ 
at the
thermal relic cross section, we expect a 5$\sigma$ significance
detection of any subhalo with
$J \sim 10^{17.2}~{\rm GeV}^2\cm^{-5}$ within  8 kpc \cite{charles}. If $J$ factors scale as the square of
the distance to the target, one could detect subhalos with $J \sim 10^{19.4}~{\rm GeV}^2\cm^{-5}$ out to 100 kpc. 
From our predicted distance of the perturbing subhalo, we
find a wide distribution that would cover that detection range. 

\begin{figure}[!t]
\centering
\includegraphics[height=7.1cm]{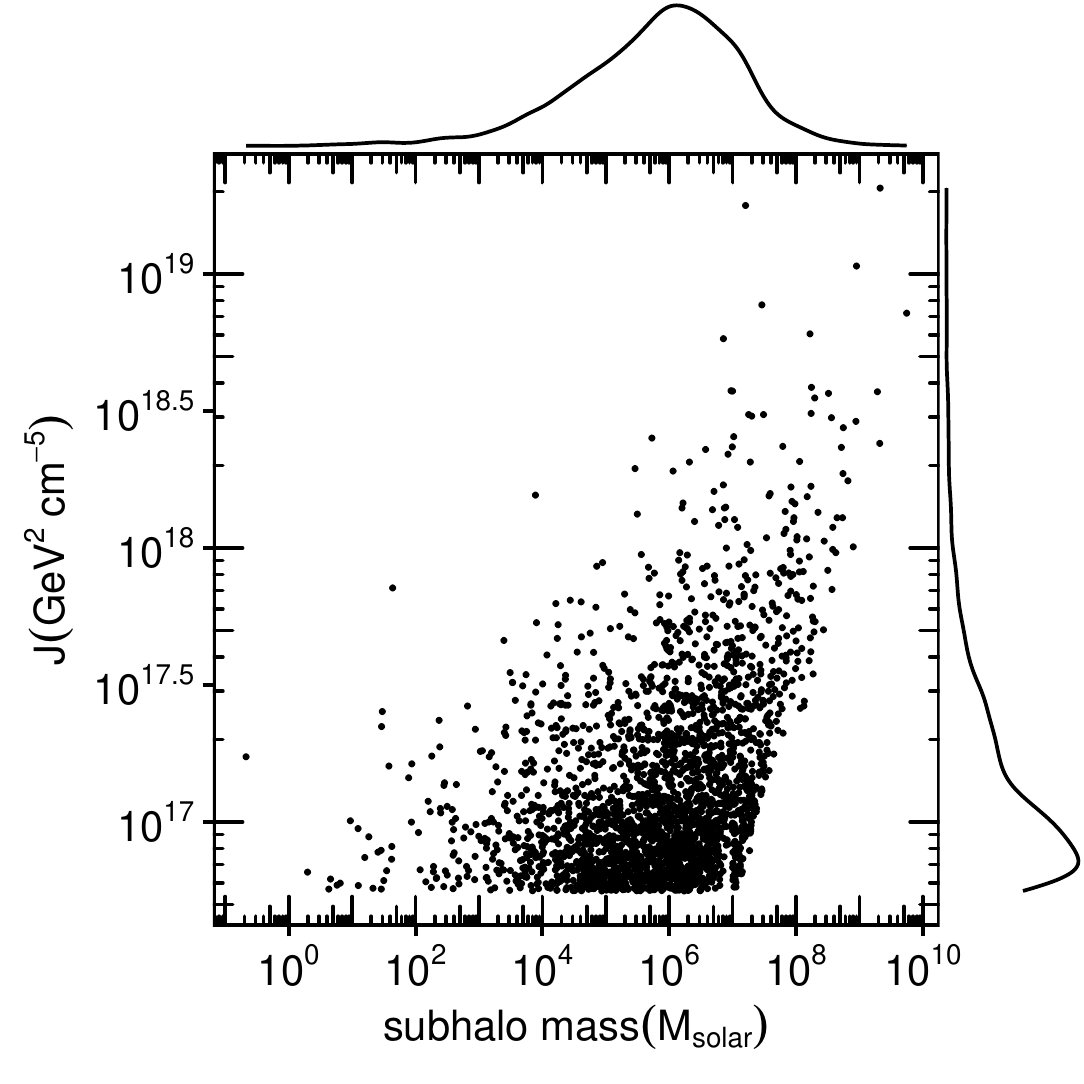}
\includegraphics[height=7.1cm]{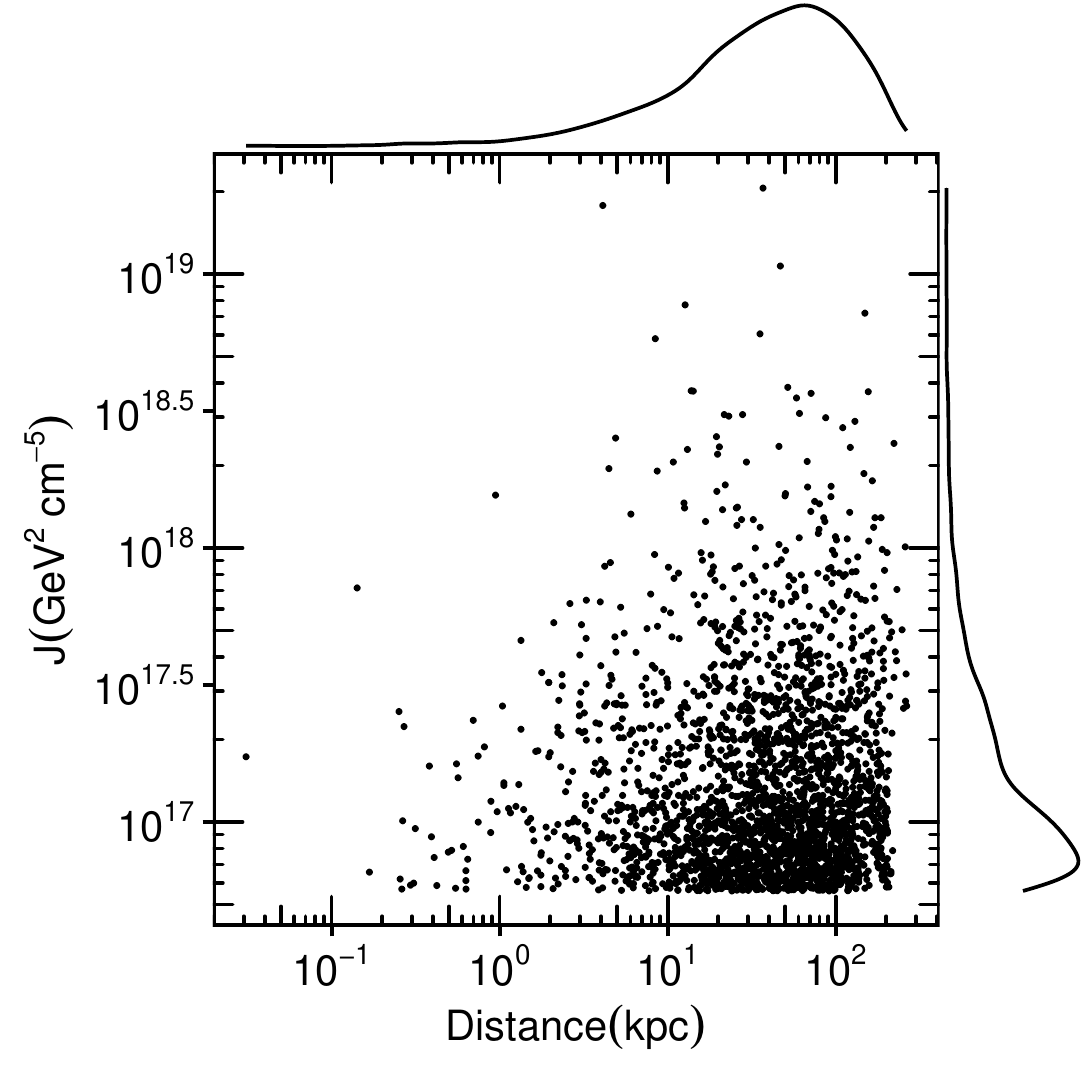}
\caption{Left: Subhalo $J$ factor as a function of subhalo mass. The points correspond to the 2500 brightest subhalos in 5 {\tt CLUMPY} realizations. 
Right: Scatter plot of subhalo $J$ factors as a function of distance from Earth for the same sample. Marginal density plot are also shown.
}
\label{fig4}
\end{figure}

To substantiate these estimates, we  use the {\tt CLUMPY} code\footnote{https://clumpy.gitlab.io/CLUMPY/} to reproduce a scaled down version of  $J$-factor predictions here. We follow
a nearly identical approach as the one used in \cite{hutten}.  We generate subhalo sets ranging from $m_{\rm min}=10^{-6} {\rm M_\odot}$  to $m_{\rm max\
} = 0.01\,{M_{\rm tot}}$ using the default value for {\tt CLUMPY} parameters. We  assume a halo mass distribution $\ensuremath{\mathrm{d}} N/\ensuremath{\mathrm{d}} m \propto m^{-\ensuremath{\alpha_m}}$ with $\ensuremath{\alpha_m}=1.9$ and set a limit of 150 subhalos between $10^8 {\rm M_\odot}$ and $10^{10} {\rm M_\odot}$. An Einasto profile is the default configuration for the subhalo profile. In Figure \ref{fig4}, we show the brightest 2500 subhalos in 5 realizations at high Galactic latitude ($|b| \geq 10\degree$). As expected, we find some of the largest J-factors in the  $\sim 10^6-10^8$ M$_\odot$ subhalo mass range within distances between 1 and 100 kpc. For a complete analysis using 500 simulations and a more complete discussion of {\tt CLUMPY} caveats see \cite{hutten}. 

\section{They Might Be Pulsars}
Throughout our discussion, we have noted that it might not be possible
to distinguish between subhalos from annihilating $\sim$ 20--70 GeV dark
matter particles and gamma-ray pulsars using spectral information alone.
Based on population syntheses of millisecond pulsars, some models predict that {\it Fermi} could detect around
170 MSPs within	10 years \cite{gonthier}. As of this writing, there are 118 MSPs  in the
Public List of LAT-Detected Gamma-Ray Pulsars \footnote{https://confluence.slac.stanford.edu/display/GLAMCOG/Public+List+of+LAT-Detected+Gamma-Ray+Pulsars}.
The upcoming Third {\it Fermi} LAT Pulsar	Catalog	(3PC in preparation) \cite{3pc} will significantly expand the census of gamma-ray pulsars and facilitate a proper comparison with the list of candidates presented here.

\section{Conclusions and Future Work}
We present a list of 73 4FGL-DR2 sources that appear to be promising dark matter subhalo candidates.
Out of the 69 covered by {\it Swift}, 17 show at least one X-ray source within the LAT ellipse.
If we only consider gamma-ray sources detected in more than one catalog (1FGL, 2FGL, 3FGL, 4FGL, 4FGL-DR2), one can narrow the list further to 40 sources.
From the analysis presented here and recurrent selection as subhalo candidates in previous works \cite{Mirabal2016,Coronado_Blazquez2019,Coronado2}, we highlight a subset of seven sources that should be further investigated as potential subhalo candidates: 4FGL J0545.7+6016, 4FGL J0953.6-1509, 4FGL J1225.9+2951, 4FGL J1539.4-3323, 4FGL J1543.6-0244, 4FGL J2112.5-3043, and 4FGL J2133.1-6432.
Out of these seven candidates, 4FGL J1539.4-3323 and 4FGL J2112.5-3043 lie in sky locations with the highest density of allowed GD-1 perturber models with predicted distances smaller than 40\,kpc.

It is important to note that the presumed subhalo
spectra from annihilating $\sim$ 20--70 GeV dark
matter particles is consistent with the intriguing Galactic Center signal \cite{goodenough,latcenter}.  
Now there are two general caveats when interpreting our results. First,
any correspondence between simulations and 4FGL-DR2 subhalo locations as in Figure~\ref{fig3}
should not be taken directly as a distance proxy. Second,
the mere presence of X-ray point sources within a
LAT 95\% error ellipse does not necessarily translate into
an actual source association \cite{4FGL}. However, X-ray sources can provide initial follow-up targets for
optical spectroscopy.

As the characterization of stellar streams improves, it might be possible to 
actually pinpoint the exact location of dark matter perturbing subhalos 
directly. This will further narrow down
the potential gamma-ray counterpart. This work is 
a first attempt at what might be possible to achieve 
in the future.

\acknowledgments
The material is based upon work supported by NASA under award number 80GSFC21M0002.
This research has made use of data obtained through the High Energy Astrophysics Science Archive Research Center Online Service, provided by the NASA/Goddard Space Flight Center.
This work made use of data supplied by the UK Swift Science Data Centre at the University of Leicester. We acknowledge valuable conversations with Javier Coronado-Bl\'azquez. We thank 
Miguel \'Angel S\'anchez-Conde for providing many helpful comments on the entire manuscript. We also thank the referee for providing helpful comments. AB acknowledges support from NASA through HST grant HST-GO-15930.

\bibliographystyle{JHEP.bst}
\bibliography{References.bib}

\clearpage
\begin{longtable}{cccccc}
    \small
    Source name & RA & Dec & P(RF) & P(XGboost) & Initial Detection\\
    \hline
    4FGL J0003.6+3059 & 0.9045 & 30.9898 & 0.88 & 0.94 & 3FGL\\
    \hline
    4FGL J0048.6$-$6347 & 12.1685 & -63.7914 & 0.88 & 0.91 & 2FGL\\
    \hline
    4FGL J0139.5$-$2228 & 24.8971 & -22.4777 & 0.54 & 0.66 & 4FGL\\
    \hline
      4FGL J0336.0+7502   & 54.0249 & 75.05 & 0.99 & 0.99 & 1FGL\\
    \hline 
    4FGL J0341.9+3153c  & 55.4852 & 31.8952 & 0.85 & 0.99 & 2FGL\\
    \hline
    4FGL J0414.7$-$4300  &  63.6977 & -43.012 & 0.90   & 0.96 & 4FGL\\
    \hline
    4FGL J0418.9+6636  & 64.7254 & 66.6001 & 1.00 &  1.00 & 2FGL\\
    \hline
    4FGL J0436.9+2915  & 69.2464 & 29.2565 & 0.72  & 0.63 & 4FGL\\
    \hline
    4FGL J0447.2+2446  & 71.8014 & 24.7701 & 0.98 & 0.97 & 4FGL\\
    \hline
    4FGL J0533.6+5945 &  83.4175 & 59.7622 & 0.98 & 0.97 & 3FGL\\
    \hline
    4FGL J0545.7+6016 & 86.4419 & 60.2704 & 0.95 & 0.99 & 1FGL\\
    \hline
    4FGL J0802.1$-$5612 & 120.5456 & -56.2012 & 1.00 & 0.99 & 1FGL\\
    \hline
    4FGL J0906.8$-$2122 & 136.7046 & -21.3724 & 0.66 & 0.84 & 4FGL\\
    \hline
    4FGL J0940.3$-$7610 & 145.0989 & -76.1794 & 0.99 & 0.96 & 1FGL\\
    \hline
    4FGL J0953.6$-$1509 & 148.4055 & -15.1549 & 0.99 & 0.99 & 1FGL\\
    \hline
    4FGL J1106.7$-$1742 & 166.699 & -17.7148 & 0.83 & 0.68 & 1FGL\\
    \hline
    4FGL J1120.0$-$2204 & 170.0016 & -22.0779 & 0.99 & 0.96 & 1FGL\\
    \hline
    4FGL J1126.0$-$5007 & 171.5145 & -50.1194 & 0.74 & 0.69 & 3FGL\\
    \hline
    4FGL J1204.5$-$5032 & 181.1483 & -50.5456 & 0.98 & 0.91 & 4FGL\\
    \hline
    4FGL J1207.4$-$4536 & 181.8734 & -45.6125 & 0.99 & 0.88 & 3FGL\\
    \hline
    4FGL J1225.9+2951 & 186.4902 & 29.8596 &  1.00 & 0.97 & 1FGL\\
    \hline
    4FGL J1231.6$-$5116 & 187.9102 & -51.2672 & 0.99 & 0.99 & 1FGL\\
    \hline
    4FGL J1345.9$-$2612 & 206.4815 & -26.2116 & 0.87 & 0.63 & 2FGL\\
    \hline
    4FGL J1400.0$-$2415 & 210.0206 & -24.266 & 0.94 & 0.99 & 2FGL\\
    \hline
    4FGL J1429.8$-$0739 & 217.4512 & -7.6506 & 0.52 & 0.69 & 4FGL\\
    \hline
    4FGL J1458.8$-$2120   & 224.7033 & -21.3388 & 0.85 & 0.99 & 2FGL\\
    \hline
    4FGL J1526.6$-$2743   & 231.6709 & -27.7327 & 1.00 & 0.98 & 4FGL\\
    \hline
    4FGL J1526.6$-$3810   & 231.6594 & -38.169 & 0.99 & 0.99 & 3FGL\\
    \hline
    4FGL J1527.8+1013     & 231.9709 & 10.2286 & 0.91 & 0.93 & 4FGL\\
    \hline
    4FGL J1530.0$-$1522     & 232.5011 & -15.3759 & 0.96 & 0.98 & 4FGL\\
    \hline
    4FGL J1539.4$-$3323     & 234.8511 & -33.3987 & 0.85 & 0.99 & 1FGL\\
    \hline
    4FGL J1543.6$-$0244     & 235.9077 & -2.7471 & 0.96 & 0.56 & 1FGL\\
    \hline
    4FGL J1544.2$-$2554     & 236.0523 & -25.9125 & 1.00 & 0.98 & 2FGL\\
    \hline
    4FGL J1602.2+2305     & 240.5576 & 23.0969 & 0.93 & 0.99 & 2FGL\\
    \hline
    4FGL J1612.1+1407     & 243.0313 & 14.1168 & 0.90 & 0.99 & 2FGL\\
    \hline
    4FGL J1622.2$-$7202     & 245.5525 & -72.0399 & 0.89 & 0.99 & 4FGL\\
    \hline
    4FGL J1623.9$-$6936 & 245.9865 & -69.6069 & 0.81 & 0.87 & 4FGL\\
    \hline
    4FGL J1630.1$-$1049 & 247.5289 & -10.8183 & 0.99 & 0.74 & 1FGL\\
    \hline
    4FGL J1646.7$-$2154 & 251.6851 & -21.9075 & 0.97 & 0.98 & 4FGL\\
    \hline
    4FGL J1656.4$-$0410 & 254.1194 & -4.1702 & 0.78 & 0.99 & 4FGL\\
    \hline
    4FGL J1659.0$-$0140 & 254.7655 & -1.6775 & 0.75 & 0.99 & 2FGL\\
    \hline
    4FGL J1700.0$-$0122 & 255.0191 & -1.3734 & 0.82 & 0.70 & 4FGL\\
    \hline
    4FGL J1709.9$-$0900 & 257.4834 & -9.0144 & 0.96 & 0.99 & 4FGL\\
    \hline
    4FGL J1711.9$-$1922 & 257.9776 & -19.3676 & 0.96 & 0.99 & 4FGL\\
    \hline
    4FGL J1717.5$-$5804 & 259.3784 & -58.0706 & 0.98 & 0.92 & 2FGL\\
    \hline
    4FGL J1720.6+0708 & 260.1638 & 7.1469 & 0.99 & 0.65 & 1FGL\\
    \hline
    4FGL J1722.8$-$0418 & 260.7115 & -4.3033 & 0.72 & 0.74 & 1FGL\\
    \hline
    4FGL J1730.4$-$0359 & 262.6086 & -3.9923 & 1.00 & 0.99 & 1FGL\\
    \hline
    4FGL J1757.7$-$6032 & 269.4489 & -60.5374 & 0.88 & 0.96 & 2FGL\\
    \hline
    4FGL J1813.5+2819 & 273.3922 & 28.3263 & 0.56 & 0.80 & 3FGL\\
    \hline
    4FGL J1818.6+1316 & 274.6527 & 13.2731 & 0.94 & 0.65 & 3FGL\\
    \hline
    4FGL J1823.2+1209 & 275.8120 & 12.1573 & 0.97 & 0.99 & 4FGL\\
    \hline
    4FGL J1824.2$-$5427 & 276.0704 & -54.4514 & 0.91 & 0.73 & 3FGL\\
    \hline
    4FGL J1827.5+1141 & 276.8786 & 11.6863  & 0.99 & 0.98 & 2FGL\\
    \hline
    4FGL J1831.1$-$6503 & 277.7773 & -65.0659 & 0.99 & 0.99 & 1FGL\\
    \hline
    4FGL J1842.1+2737 & 280.5378 & 27.6248 & 0.93 & 0.61 &2FGL\\
    \hline
    4FGL J1845.8$-$2521 & 281.4648 & -25.3585 & 0.99 & 0.78 & 3FGL\\
    \hline
    4FGL J1855.6$-$3603 & 283.9149 & -36.0611 & 0.97 & 0.78 & 4FGL\\
    \hline 
    4FGL J1858.3$-$5424 & 284.576 & -54.4123 & 0.91 & 0.82 & 3FGL\\
    \hline
    4FGL J1906.4$-$1757 & 286.6107 & -17.9509 & 0.71 & 0.99 & 4FGL\\
    \hline
    4FGL J1910.7$-$5320 & 287.6989 & -53.3385 & 0.83 & 0.94 & 1FGL\\
    \hline
    4FGL J1913.4$-$1526 & 288.3515 & -15.4496 & 0.97 & 0.85 & 4FGL\\
    \hline
    4FGL J1920.0$-$2622 & 290.023 & -26.3775 & 0.52 & 0.52 & 4FGL\\
    \hline
    4FGL J1924.8$-$1035 & 291.2053 & -10.5909 & 1.00 & 1.00 & 2FGL\\
    \hline
    4FGL J1949.2$-$1453 & 297.3162 & -14.8983 & 0.74 & 0.54 & 2FGL\\
    \hline
    4FGL J2026.3+1431 & 306.5988 & 14.5225 & 0.74 & 0.79 & 3FGL\\
    \hline
    4FGL J2029.5$-$4237 & 307.3765 & -42.6285 & 0.91 & 0.74 & 3FGL\\
    \hline
    4FGL J2043.9$-$4802 & 310.9815 & -48.0398 & 0.95 & 0.98 & 2FGL\\
    \hline
    4FGL J2112.5$-$3043 & 318.1400 & -30.7293 & 0.96 & 0.99 & 1FGL\\
    \hline
    4FGL J2133.1$-$6432 & 323.2951 & -64.5383 & 1.00 & 0.99 & 1FGL\\
    \hline
    4FGL J2212.4+0708 & 333.1083 & 7.1428 & 0.99 & 0.92 & 1FGL\\
    \hline
    4FGL J2219.7$-$6837 & 334.9469 & -68.6173 & 0.61 & 0.93 & 3FGL\\
    \hline
    4FGL J2250.5+3305 & 342.6404 & 33.0989 & 0.97 & 0.99 & 3FGL\\
    \hline
    \caption{Subhalo candidates in the 4FGL-DR2\\
}
    \label{table1}
\end{longtable}
\clearpage
\begin{longtable}{ccccc}
    \small
    Source name & $N_{\rm H}$ (cm$^{2}$) & X-ray Flux Upper Limit (erg cm$^{-\
2}$ s$^{-1}$)\\
    \hline
    \hline
    4FGL J0003.6+3059 & $4.9 \times 10^{20}$ & $1.5 \times 10^{-13}$(PS)\\
    \hline 
    4FGL J0048.6$-$6347 &$2.2 \times 10^{20}$ & $ < 8.4 \times 10^{-14}$\\
    \hline 
    4FGL J0139.5$-$2228 &$9.9 \times 10^{19}$ & $ < 8.9 \times 10^{-14}$\\
    \hline 
    4FGL J0336.0+7502 & $1.5 \times 10^{21}$ & $4.5 \times 10^{-14}$(PS)\\
    \hline 
    4FGL J0341.9+3153c & $1.2 \times 10^{21}$ & $8.6 \times 10^{-13}$(PS)\\
    \hline
    4FGL J0436.9+2915 & $1.7 \times 10^{21}$ & $ < 1.6 \times 10^{-13}$\\
    \hline
    4FGL J0447.2+2446 & $2.1 \times 10^{21}$ & $ < 8.6 \times 10^{-13}$\\
    \hline
    4FGL J0533.6+5945 & $1.8 \times 10^{21}$ & $ < 1.2 \times 10^{-13}$\\
    \hline
    4FGL J0545.7+6016 & $1.5 \times 10^{21}$ &$ < 2.0 \times 10^{-13}$ \\
    \hline
    4FGL J0802.1$-$5612 &  $1.4 \times 10^{21}$ & $ 4.1 \times 10^{-14}$(PS)\\
    \hline
    4FGL J0906.8$-$2122 & $1.1 \times 10^{21}$ &  $ < 8.1 \times 10^{-14}$\\
    \hline
    4FGL J0940.3$-$7610 & $9.6 \times 10^{20}$ & $2.2 \times 10^{-13}$(PS)\\
    \hline
    4FGL J0953.6$-$1509 & $5.8 \times 10^{20}$ & $ < 7.7 \times 10^{-14}$\\
    \hline
    4FGL J1106.7$-$1742 & $4.0 \times 10^{20}$ & Partially Covered\\
    \hline
    4FGL J1120.0$-$2204 & $4.1 \times 10^{20}$ & $ 7.2 \times 10^{-14}$(PS)\\
    \hline
    4FGL J1126.0$-$5007 & $9.1 \times 10^{20}$ & $ 6.2 \times 10^{-14}$(PS) 
(also GRB 140719A in error ellipse)\\
    \hline
    4FGL J1204.5$-$5032 & $1.1 \times 10^{21}$ & $ < 7.5 \times 10^{-14}$\\
    \hline
    4FGL J1207.4$-$4536 &$6.7 \times 10^{20}$  & $2.8 \times 10^{-13}$(PS)\\
    \hline
    4FGL J1225.9+2951 &$1.6 \times 10^{20}$ & $ < 8.7 \times 10^{-14}$\\
    \hline
    4FGL J1231.6$-$5116 &  $1.4 \times 10^{21}$ & $ < 1.2 \times 10^{-13}$\\
    \hline
    4FGL J1345.9$-$2612 & $4.7 \times 10^{20}$ & $ < 8.6 \times 10^{-14}$ \\
    \hline
    4FGL J1400.0$-$2415 & $5.1 \times 10^{20}$ & $ < 6.4 \times 10^{-14}$\\
    \hline
    4FGL J1429.8$-$0739 & $4.6 \times 10^{20}$ & $ 8.5 \times 10^{-13}$(PS)\\
    \hline
    4FGL J1458.8$-$2120 & $7.4 \times 10^{20}$ & $ < 8.3 \times 10^{-14}$\\
    \hline
    4FGL J1526.6$-$3810 & $1.0 \times 10^{21}$ & $ < 1.4 \times 10^{-13}$\\
    \hline
    4FGL J1526.6$-$2743 & $7.7 \times 10^{20}$ & $ < 1.4 \times 10^{-13}$ \\
    \hline
    4FGL J1527.8+1013 & $2.8 \times 10^{20}$ & $ 1.1 \times 10^{-13}$(PS)\\
    \hline
    4FGL J1530.0$-$1522 & $8.3 \times 10^{20}$ & $ 1.4 \times 10^{-13}$(PS)\\
    \hline
    4FGL J1539.4$-$3323 & $9.0 \times 10^{20}$ & $< 9.0 \times 10^{-15}$\\
    \hline
    4FGL J1543.6$-$0244  &$7.9 \times 10^{20}$ &$ < 9.6 \times 10^{-14}$\\
    \hline
    4FGL J1544.2$-$2554 &$1.1 \times 10^{21}$ & $ < 9.9 \times 10^{-14}$\\
    \hline 
    4FGL J1602.2+2305 &$4.3 \times 10^{20}$ & $ < 1.2 \times 10^{-13}$\\
    \hline
    4FGL J1612.1+1407 & $3.1 \times 10^{20}$ & $< 8.4 \times 10^{-14}$\\
    \hline
    4FGL J1622.2$-$7202 & $7.5 \times 10^{20}$ & $ 8.9 \times 10^{-14}$(PS)\\
    \hline
    4FGL J1623.9$-$6936 & $9.7 \times 10^{20}$ &$ < 7.7 \times 10^{-14}$\\
    \hline
    4FGL J1627.7$-$5749 & $2.7 \times 10^{21}$ & Partially Covered\\
    \hline
    4FGL J1630.1$-$1049 & $1.3 \times 10^{21}$ & $ < 1.1 \times 10^{-13}$\\
    \hline
    4FGL J1646.7$-$2154 &$1.3 \times 10^{21}$ & Not Covered\\
    \hline
    4FGL J1656.4$-$0410  &$1.0 \times 10^{21}$ & $ 1.2 \times 10^{-13}$(PS)\\
    \hline
    4FGL J1659.0$-$0140 & $8.4 \times 10^{20}$ & $ < 1.3 \times 10^{-13}$\\
    \hline
    4FGL J1700.0$-$0122 & $8.5 \times 10^{20}$ & $ 2.6 \times 10^{-13}$(PS)\\
    \hline
    4FGL J1709.9$-$0900 & $1.5 \times 10^{21}$ & $ < 1.1 \times 10^{-13}$\\
    \hline
    4FGL J1711.9$-$1922 &$1.5 \times 10^{21}$ &$ < 1.9 \times 10^{-13}$ \\
    \hline
    4FGL J1717.5$-$5804 & $9.2 \times 10^{20}$ &$ < 2.6 \times 10^{-13}$ \\
    \hline
    4FGL J1720.6+0708 & $6.2 \times 10^{20}$ &$ < 1.0 \times 10^{-13}$ \\
    \hline
    4FGL J1722.8$-$0418 & $1.3 \times 10^{21}$ & $ < 1.0 \times 10^{-13}$\\
    \hline
    4FGL J1730.4$-$0359 & $1.5 \times 10^{21}$ & $ < 9.0 \times 10^{-13}$\\
    \hline
    4FGL J1757.7$-$6032 & $6.8 \times 10^{20}$ & $ < 1.2 \times 10^{-13}$\\
    \hline
    4FGL J1813.5+2819 &$7.4 \times 10^{20}$ & $ < 1.2 \times 10^{-13}$\\
    \hline
    4FGL J1818.6+1316 &$9.8 \times 10^{20}$ & $ < 1.2 \times 10^{-13}$\\
    \hline
    4FGL J1823.2+1209 & $1.3 \times 10^{21}$ & $ < 1.3 \times 10^{-13}$\\
    \hline
    4FGL J1824.2$-$5427 &$7.7 \times 10^{20}$ & $ < 1.3 \times 10^{-13}$\\
    \hline
    4FGL J1827.5+1141 &$1.4 \times 10^{21}$ & $ < 1.1 \times 10^{-13}$\\
    \hline
    4FGL J1831.1$-$6503 &$6.1 \times 10^{20}$ & $ < 1.1 \times 10^{-13}$\\
    \hline
    4FGL J1842.1+2737 &$1.1 \times 10^{21}$ & $ < 1.1 \times 10^{-13}$\\
    \hline
    4FGL J1845.8$-$2521 &$1.2 \times 10^{21}$ & $ < 1.3 \times 10^{-13}$\\
    \hline
    4FGL J1855.6$-$3603 & $6.4 \times 10^{20}$ & $ < 1.4 \times 10^{-13}$\\
    \hline
    4FGL J1858.3$-$5424 & $6.0 \times 10^{20}$ &$ < 1.8 \times 10^{-13}$\\
    \hline
    4FGL J1906.4$-$1757 &$9.1 \times 10^{20}$ & Not covered\\
    \hline
    4FGL J1910.7$-$5320 &$5.3 \times 10^{20}$ & $ < 3.3 \times 10^{-13}$\\
    \hline
    4FGL J1920.0$-$2622 &$7.9 \times 10^{20}$ & $ < 1.2 \times 10^{-13}$\\
    \hline
    4FGL J1924.8$-$1035 &$1.2 \times 10^{21}$ &$ < 1.1 \times 10^{-13}$\\
    \hline    
    4FGL J1949.2$-$1453&$7.8 \times 10^{20}$ &< $1.1 \times 10^{-13}$\\
    \hline
    4FGL J2026.3+1431 & $6.2 \times 10^{20}$     & < $1.0 \times 10^{-13}$\\
    \hline
    4FGL J2029.5$-$4237 &$3.7 \times 10^{20}$ & $ 3.7 \times 10^{-13}$(PS)\\
    \hline
    4FGL J2043.9$-$4802 & $2.7 \times 10^{20}$ & $< 5.2 \times 10^{-14}$ \\
    \hline
    4FGL J2112.5$-$3043 & $6.4 \times 10^{20}$ & $< 2.5 \times 10^{-14}$\\
    \hline
    4FGL J2133.1$-$6432 &$3.2 \times 10^{20}$ & $< 6.0 \times 10^{-14}$\\
    \hline
    4FGL J2212.4+0708 & $6.2 \times 10^{20}$ &$1.3 \times 10^{-13}$(PS)\\
    \hline
    4FGL J2219.7$-$6837&$2.8 \times 10^{20}$ & < $1.5 \times 10^{-13}$\\ 
    \hline
    4FGL J2250.5+3305&$8.0 \times 10^{20}$ & $ 2.0 \times 10^{-13}$(PS)\\
    \hline
    \caption{ X-ray upper limits for 4FGL-DR2 subhalo candidates. PS denotes at least one X-ray point source
within the LAT error ellipse. Partial or no {\it Swift} coverage are also  indicated.
}
    \label{table2}
\end{longtable}

\end{document}